# Development and testing of cost-effective, 6 cm × 6 cm MCP-based photodetectors for fast timing applications

Jingbo Wang[a]*, Karen Byrum[a], Marcel Demarteau[a], Jeffrey Elam[b], Anil Mane[b], Edward May[a], Robert Wagner[a], Dean Walters[a], Junqi Xie[a], Lei Xia[a], Huyue Zhao[a]

[a]*High Energy Physics Division, Argonne National Laboratory, Argonne, IL 60439, USA*
[b]*Energy Systems Division, Argonne National Laboratory, Argonne, IL 60439, USA*

**Abstract:** Micro-channel plate (MCP)-based photodetectors are capable of picosecond level time resolution and sub-mm level position resolution, which makes them a perfect candidate for the next generation large area photodetectors. The large-area picosecond photodetector (LAPPD) collaboration is developing new techniques for making large-area photodetectors based on new MCP fabrication and functionalization methods. A small single tube processing system (SmSTPS) was constructed at Argonne National Laboratory (ANL) for developing scalable, cost-effective, glass-body, 6 cm × 6 cm, picosecond photodetectors based on MCPs functionalized by Atomic Layer Deposition (ALD). Recently, a number of fully processed and hermetically sealed prototypes made of MCPs with 20 μm pores have been fabricated. This is a significant milestone for the LAPPD project. These prototypes were characterized with a pulsed laser test facility. Without optimization, the prototypes have shown excellent results: The time resolution is ~57 ps for single photoelectron mode and ~15 ps for multi-photoelectron mode; the best position resolution is ≤ 0.8 mm for large pulses. In this paper, the tube processing system, the detector assembly, experimental setup, data analysis and the key performance will be presented.

## 1. Introduction

Photomultiplier tubes (PMTs) [1] are current-source amplifiers with high gain, high quantum efficiency (QE) and the ability to resolve single photons. PMTs have a time resolution of a few hundred picoseconds [2], but the complexity of their discrete dynode configuration limits their capability to provide precise spatial information and makes them susceptible to magnetic fields. Multi-anode PMTs with a more compact structure have similar performances as standard PMTs and allow for better position information as well [3]. However, both standard and multi-anode PMTs are not suitable for applications which require simultaneously very high time (<50 ps) and position resolution (<50 μm). Recently, the demands of nuclear and particle physics experiments call for the next generation of photodetectors with capabilities better than those of traditional photomultipliers [4][5]. Therefore, alternative technologies for fast single photon detectors have been developed, such as the microchannel plate photomultiplier tubes (MCP-PMTs).

MCP-PMTs [6] are compact photodetectors capable of picosecond level time resolution [7][8] and sub-mm position resolution [9][10], and are an evolution from the basic principles of traditional PMTs. As a key element, a

*corresponding author. Tel.: +1 630 252 5333
*E-mail address: wjingbo@anl.gov*

MCP consists of millions of parallel conductive glass capillaries (6 – 40 μm). Each capillary is an independent secondary-electron multiplier that can be considered as a continuous dynode structure. Such a structure provides local electron multiplication with a very small path length, resulting in an extremely fast time response. Compared to other photon sensors, such as traditional PMTs and solid-state detectors, the new generation of MCP-PMTs described here are able to simultaneously provide high-gain ($10^6$ – $10^7$), high QE (>20%), excellent time and position resolutions. In addition, MCP-PMTs have shown potential in applications involving strong magnetic fields [11][12][13]. All these characteristics make them an excellent candidate for the next generation of photodetectors. If MCP-PMTs could be made considerably more cost-effective and robust, they would be widely used in a variety of applications in the fields of particle and nuclear physics, astrophysics and medical imaging.

The Large-Area Picosecond Photodetector (LAPPD) project [14] is a US Department of Energy (DOE) funded collaborative project, with the goal of developing low cost, commercializable methods to fabricate 400 $cm^2$ thin planar photodetectors based on Atomic Layer Deposition (ALD) coated MCPs [16]. The production process of the MCP substrates, developed by Incom Inc[1], is based on the use of hollow capillaries in a glass drawing process, taking the place of the traditional chemical etching process. The MCP substrates are functionalized by an ALD coating method at ANL, which controls the necessary resistive and secondary emission properties [15]. The ability to engineer the material properties offers the opportunity to improve the performance and reduce the cost for MCP-based devices. During the past few years, the LAPPD collaboration has made substantial progress on the MCP fabrication and the photodetector development [15].

Argonne National Laboratory has built a photodetector processing system, capable of producing 6 cm × 6 cm active area, glass body, cost-effective photodetectors. Two critical issues were addressed for the production of robust photodetectors: indium-based hermetic sealing and MCP outgassing. Recently, a series of long-lived prototype photodetectors were produced, representing a significant milestone for the production of hermetically sealed functional devices. A testing facility was built for characterizing the MCP-based photodetectors. It uses a pulsed blue laser (405 nm)[2] with duration time of 70 ps FWHM and allows for testing the fast time response as well the gain and uniformity of the photodetectors. In this paper, we present the development and testing of the 6 cm × 6 cm MCP-based photodetectors.

The paper is organized as follows: Section 2 describes the design of the photodetector processing system; Section 3 presents the structure of the MCP-based photodetectors; Section 4 describes the laser test facility and the experimental setup; Section 5 describes the data analysis method; Section 6 presents the performance of the photodetectors. Finally, Section 7 gives our conclusions and outlook.

## 2. Small single tube processing system

The LAPPD collaboration initially planned to build a photodetector production facility for producing 20 cm × 20 cm MCP photodetectors in an all-glass package one detector at a time. Considering the high cost and complexity of building such a large system, ANL has built a smaller system that would serve as an R&D facility and as an intermediate step towards the production of the full-size detector and their commercialization. The small single tube processing system (SmSTPS) has the ability to produce planar, glass-body, small form factor photodetectors (6 cm × 6 cm) with high QE bialkali photocathodes. The tube processing system consists of four major parts (See Fig. 1):

1) Vacuum Load-lock: A chamber with a load-lock for loading the photodetector base, the internal stack and the top window.
2) Bake and scrub chamber: A chamber for baking and scrubbing the MCPs and the top window on which the photocathode is deposited.
3) Photocathode deposition chamber: A chamber for photocathode evaporation.
4) Sealing chamber: A chamber for making the final seal of the top window to the assembled photodetector base with the internal stack of MCPs and grid spacers.

---

[1] Incom Inc: http://www.incomusa.com/
[2] Hamamatsu PLP-10 pulsed blue laser

All the chambers are isolated by means of gate valves on either side. The system is equipped with magnetic transfer arms which enable an external control of transferring components between chambers. The device integration is based on a "transfer process" procedure. In this process, the ALD-coated MCPs and grid spacers are assembled with the photodetector base and conditioned in the scrubbing chamber. The detector base with the internal components is then moved to the sealing chamber. Meanwhile, the top window is coated with a bialkali photocathode at the bottom surface, and then transferred to the sealing chamber. The final seal relies on a low temperature thermo-compression sealing process that joins the top window to the glass-body stack. A more detailed description can be found in Ref. [23]. The SmSTPS is expected to produce one detector per week with a yield higher than 50%. It is worth noting that this facility is presently serving as an R&D and production facility at the same time. The advantage of this system is its flexibility, which enables the development of various detector configurations and as well as fabrication methods. The experience can then be transferred to industry for large scale production.

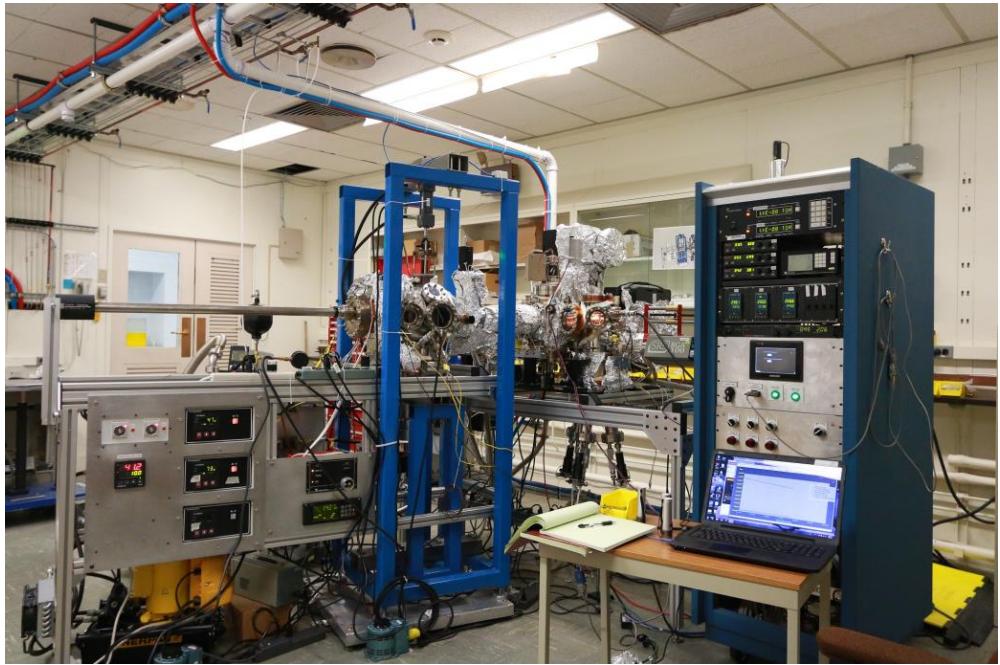

Fig 1. Picture of the small single tube processing system at ANL.

## 3. Design of the 6 cm × 6 cm MCP-based photodetector

*3.1 Stack components*

The modular design of the MCP-PMT is shown in Fig. 2. This design inherits its design concept from the LAPPD detector [26] but with a small form factor, based on an all-glass hermetically sealed tube package that consists of the following components:
1) A glass bottom plate on which is silk-screened a stripline anode readout [25].
2) A glass side wall.
3) A resistively matched pair of MCPs separated by a grid spacer.
4) Three glass grid spacers.
5) A glass top window with a bialkali (K, Cs) photocathode.
6) An indium seal between the top window and the sidewall.

The side wall is glass-frit bonded to the bottom plate producing what is termed the Photodetector Base. The anode readout strips extend through the side wall from vacuum to air. As the key element, the MCPs are made of microchannel glass substrates [15][27] coated by ALD [16][17][18], featuring 20 μm pores and a length-to-diameter (L/D) ratio of 60. The open area ratio is about 60%. These plates, functionalized with a resistive layer and a secondary emission layer using ALD, are arranged in a "Chevron" configuration (double plates) with 8° pore bias angles, providing electron multiplication with a gain as high as $10^7$. The grid spacers are located between the top window and the upper MCP, between the two MCPs, and between the lower MCP and the bottom anode plate. The grid spacers are also coated with a resistive layer by ALD. In addition to supporting the sealed, evacuated photodetector from implosion due to the atmospheric pressure load, they produce the desired HV biasing of the MCPs. This arrangement can then be biased by a single HV connection outside the vacuum. The glass top window has a Nichrome border providing contact between the photocathode surface and the external bias HV. A bialkali photocathode is evaporated onto the inner region of the top window.

All the package components are made of borosilicate float glass that is considerably less expensive than the leaded glass of typical commercial MCPs. The top window, the square side wall and the bottom window make a hermetically sealed package. First, the side wall is hermetically bonded in air to the bottom anode plate using a glass frit seal, producing the photodetector base. Then the internal components, such as the MCPs, grid spacers and getter strips (not shown), are positioned inside the photodetector base. This assembly and the top window is loaded into the vacuum system and processed. In the last step, the top window and the photodetector base are joined together through a low temperature thermo-compression indium seal. This seal must be made at a relatively low temperature to avoid damaging the bialkali photocathode. The nominal specifications of the stack components are summarized in Table 1.

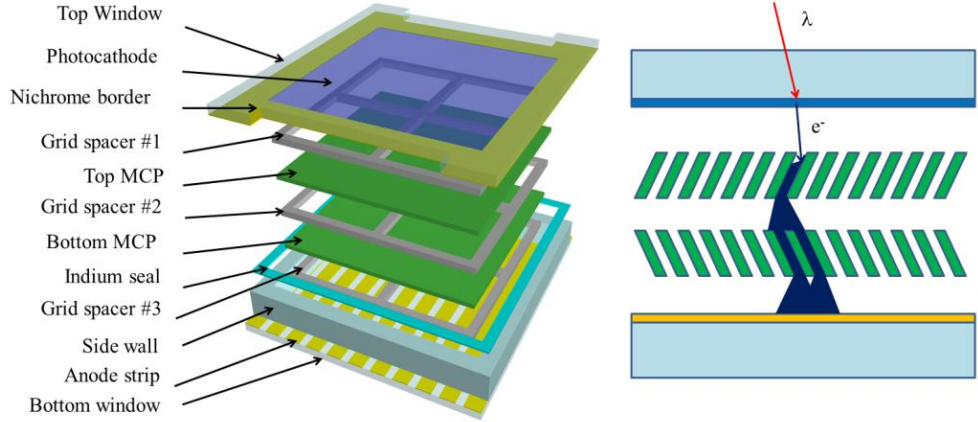

Fig 2. Construction of the 6 cm × 6 cm photodetector. The hermetic seal is provided by the top window, the side wall and the bottom window. The side wall and the bottom window are bonded together by a glass frit seal, serving as the base package. An indium seal is made between the top window and the side wall, through a low temperature thermo-compression sealing technique.

*3.2 HV divider*

In this design, all the internal stack components are resistively coated by ALD. The three grid spacers and two MCPs form a resistor chain that serves as a HV divider [21]. A negative HV is applied on the photocathode through the Nichrome border, providing a continuous DC current across the whole stack. The individual HV for internal components is defined by the voltage drop through the resistive components. Such a resistor chain HV design has no pins penetrating the glass package, which is unique in providing a simplified hermetic fabrication. However, it requires a highly matched resistance for each internal stack component. The resistor chain design is currently our default solution. So far, most of the photodetectors have been built with this design and we have achieved reliable sealing and very good detector performance.

Table 1 Specifications of the stack components

| Components | Thickness (mm) | Size (mm) | Resistive coating | Desirable resistance (MΩ) |
|---|---|---|---|---|
| Top Window | 2.75 | 85.3 × 85.3 | Bialkali | $10^5 - 10^7$ |
| Grid spacer #1 | 2 | 59.7 × 59.7 | ALD-Spacer | 40 |
| Top MCP | 1.2 | 59.7 × 59.7 | ALD-MCP | 200 |
| Grid spacer #2 | 2 | 59.7 × 59.7 | ALD-Spacer | 40 |
| Bottom MCP | 1.2 | 59.7 × 59.7 | ALD-MCP | 200 |
| Grid spacer #3 | 3.15 | 59.7 × 59.7 | ALD-Spacer | 60 |
| Side wall | 9 | 76.2 × 76.2 | - | $10^5 - 10^7$ |
| Bottom window | 2.75 | 85.3 × 85.3 | - | $10^5 - 10^7$ |

*3.3 Readout circuit board*

For signal readout, a printed circuit board was designed and integrated as part of the detector package. The circuit board has 9 traces on either side of the photodetector, matching the 9 anode strips. Connection between the fanout traces and the anode strips are made by soldering a wire. The signal trace maintains a 50 Ω characteristic impedance, thus the charge avalanche appears as a signal on a short 50 Ω transmission line for connection to an external readout/digitizing system. The DC path from signal to HV ground is provided by surface mount drain resistor of 10 kΩ. The 6 cm × 6 cm photodetector is attached to the circuit board, providing a permanent mount and solid electrical connections. Fig. 3 shows a picture of the 6 cm × 6 cm photodetector with a circuit readout board.

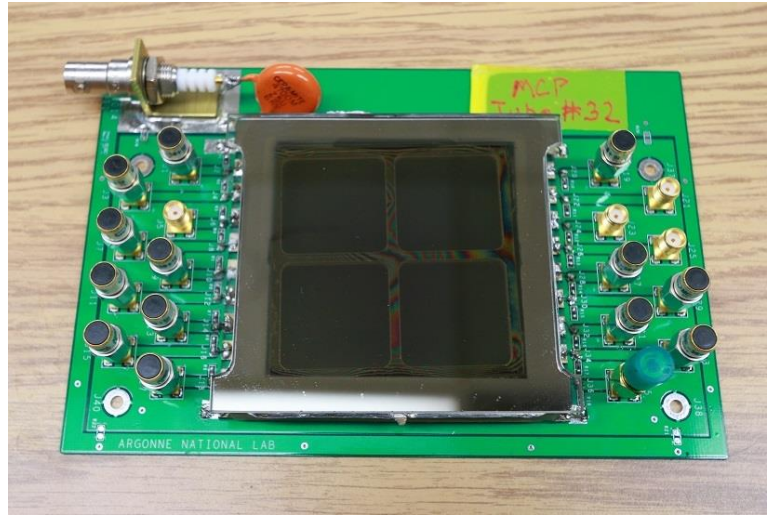

Fig. 3 Picture of the 6 cm × 6 cm photodetector. The photodetector is attached to the circuit board, providing a permanent mount and solid electrical connections.

## 4. Laser test facility at ANL

*4.1 Experimental setup*

We have established a laser based test facility for characterizing the photodetectors, using a pulsed laser emitting blue light with a wavelength of 405 nm and a typical pulse width of 70 ps FWHM. The experimental setup is shown schematically in Fig. 4. The optics are implemented in two stages that are covered by light-tight dark boxes.

Optics in the first dark box is designed to produce a narrow laser beam. The blue light emitted from the laser diode, with an emission angle of 15 degrees, is attenuated by two discrete neutral density filter wheels going through a collimator with 2 mm diameter. The light is then directed by two reflection mirrors, further collimated by an iris and fed to the second dark box. A shutter between the two boxes is used to block the laser during the test without a system shut down.

The second dark box serves as a simple testing stage for systematically characterizing the MCP photodetectors. A 50/50 splitter is located upstream next to the middle shutter, directing half the light to a reference photodiode that is used to monitor the beam variation, and half the light to the detectors under test. The light can be remotely further attenuated to the few-photon level by an additional computer controlled continuous rotating neutral density filter. A transmission level up to three orders of magnitude is achievable using three filters, which allows a wide-range scan in the number of photos. Finally, the light is split towards two test stands, allowing characterization of two detectors at the same time.

One of the test stands is attached to a two-dimensional translation stage[3] that allows 5 cm movement in both horizontal and vertical directions. The translation stage provides micrometer level precision, while maintaining a perpendicular beam incidence to the detector surface. Thus, a precise computer controlled two-dimensional position scan is feasible during illumination. The beam spot was estimated to be 1 mm, depending on the iris opening area. For measurements described in this paper, an Argonne photodetector is placed on the translation stage and a reference Photonis Planacon [28] is placed on the fixed stage, as depicted in Fig.4.

The signals from the anode strips are brought out by SMA (Sub-Miniature version A) cables. Unused anode strips are terminated by 50 Ω terminators on both ends. For data acquisition, we use a Tektronix oscilloscope (DPO7354) with a 3.5 GHz bandwidth and a maximum 40 Gs/s sampling rate. We operate the scope in fast frame mode, which allows to automatically store multiple channels of digitized waveforms under various triggers to the disk. The raw waveforms, which are scanned in a 10 ns window with 10 Gs/s sampling rate, can then be analyzed offline (see Section 5 for details).

In the current setup, if the light intensity is attenuated to a very low level, the photodiode will not be able to produce a detectable signal. As a consequence, we use the laser synchronization signal as the start time for tests at low light intensity. The photon arrival time is defined as the timing of the MCP signal relative to the start signal. Table 2 summarizes the main specifications of the characterization facility.

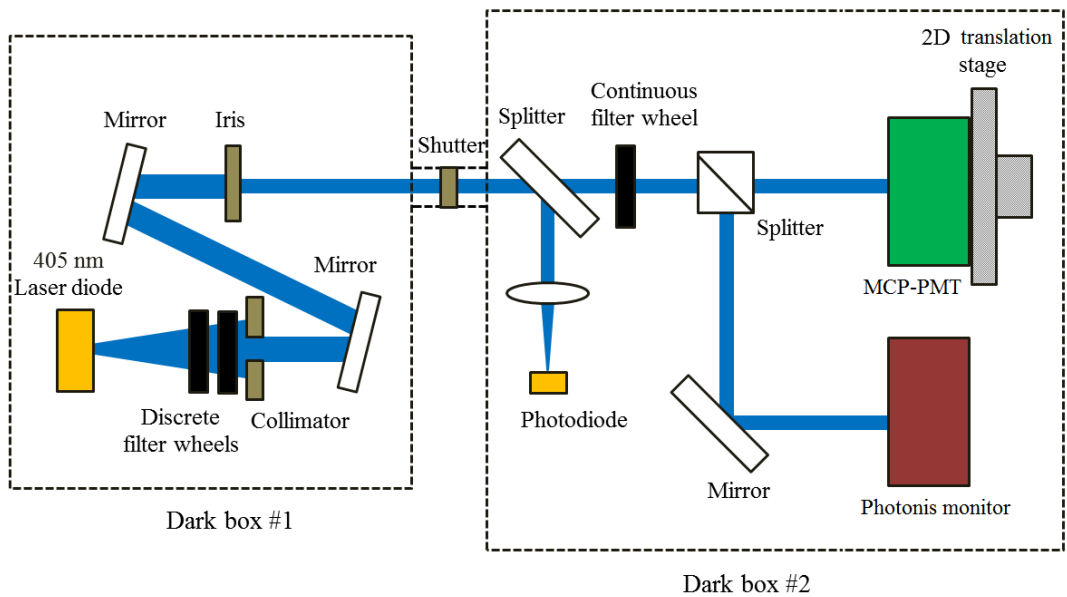

Fig. 4 Schematic of the experimental setup used for systematic characterization of the photodetectors.

---

[3] Newport Corporation: ESP300 motion controller/driver

Table 2. Specifications of the laser test facility.

| Functionality | Parameter | Capabilities |
|---|---|---|
| Laser | Pulse frequency | 2 Hz – 100 MHz |
| | Pulse duration | Typical 70 ps FWHM |
| | Wavelength | 405 nm |
| Optics | Position precision | ~ 1 μm |
| | Beam diameter | 1 – 2 mm |
| 2D translation stage | Motorized translation | 5 cm range with μm level precision |
| Oscilloscope | Model | Tektronix DPO7354 |
| | Number of channels | 4 testing channels plus one external trigger channel |
| | Noise | ~ 1 mV per sample |
| | Sampling rate | 40 Gs/s for single-channel operation, 10 Gs/s for 4-channel operation |
| | Analog bandwidth | 3.5 GHz |
| Data acquisition | - | Waveform recording through the Oscilloscope |
| Data analysis | - | Waveform processing offline in software |

*4.2 Determination of the single photoelectron light intensity*

The light intensity per pulse can be calibrated by the photodetectors assuming the number of photoelectrons follow a Poisson distribution:

$$P(N_{pe}) = \frac{\overline{N_{pe}}^{N_{pe}}}{N_{pe}!} e^{-\overline{N_{pe}}} \qquad (1)$$

where $N_{pe}$ is the number of photoelectrons, and $\overline{N_{pe}}$ is the average number of photoelectrons. Thus, the average number of photoelectrons can be calculated by:

$$\overline{N_{pe}} = -\ln(P(0)) \qquad (2)$$

where $P(0)$ is the probability of zero photoelectron. Among the valid photoelectron signals, the single photoelectron fraction $\eta$ can be calculated by the probability of zero photoelectron as:

$$\eta = \frac{P(1)}{1 - P(0)} = \frac{-P(0)\ln(P(0))}{1 - P(0)} \qquad (3)$$

Assuming a Poisson distribution, Fig. 5 shows the relation between the probability of a certain number of photoelectrons and the average number of photoelectrons. Without attenuation of the laser light, each pulse contains a large amount of photoelectrons, resulting in almost 100% observed MCP signals. By attenuating the light with the neutral density filters, we are able to reach a very low light level where more than 80% of the laser pulses produce no signal. In this case, the probability of producing more than one photoelectron is statistically suppressed and the single photoelectron fraction is high ($\eta > 90\%$). In this paper, we define the light intensity that meets a condition of $\eta>90\%$ as the single photoelectron light level. The area shaded grey in Fig. 5 indicates when the condition $\eta>90\%$ is met. Equivalently, once the average number of photoelectrons is below 0.22, we assume the single photoelectron operating mode has been reached. This is our definition of the single photoelectron mode. In this mode, the gain and timing properties are completely independent of the quantum efficiency of the photocathode. The same method was also used in our previous measurements, applying a condition of $\eta>95\%$ [20].

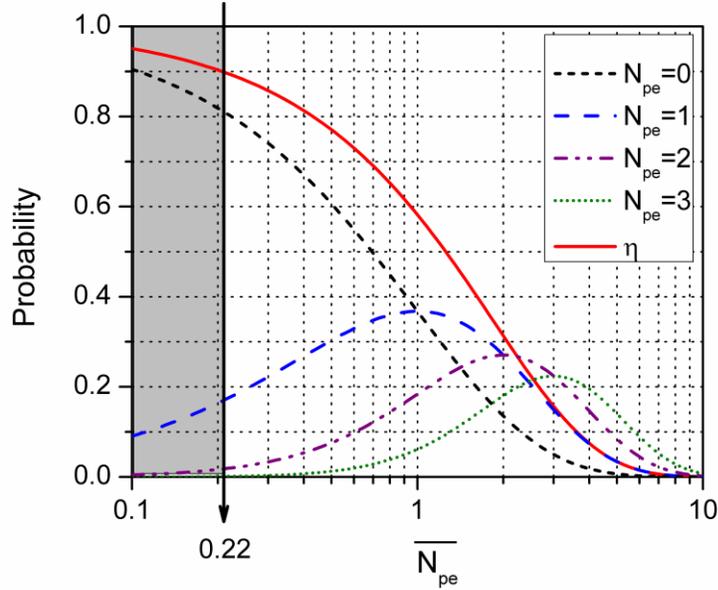

Fig. 5 Photoelectron probability as a function of the average number of photoelectrons. If the average number of photoelectrons is less than 0.22, the probability of producing more than one photoelectron is negligible. The red curve gives the single photoelectron fraction.

## 5. Data analysis

The waveforms are recorded by a Tektronix DPO7354 oscilloscope with a 40 Gs/s sampling rate and a 3.5 GHz bandwidth, and the data analysis is done offline in software. The data analysis is done in several steps, including waveform processing, pulse selection and timing discrimination.

Prior to the pulse selection, a Fast Fourier Transform (FFT) algorithm is applied to the raw waveforms. In the frequency domain, the highest signal component is about 1 GHz (see Section 6.1), well below the analog bandwidth of the oscilloscope. Frequency components higher than 1 GHz are mainly from the noise. Therefore, a low pass frequency filter is applied to suppress the high frequency noise, which improves the signal shape. This is important for single photoelectron operation where the signal to noise ratio is low.

Pulse selection is performed through a Time over Threshold (TOT) cut. A good pulse is defined as a pulse having a time duration longer than 700 ps over a 2 mV threshold. The threshold is set a bit higher than the maximum amplitude of the baseline and the minimum duration is set by the typical rise time. Once a signal is identified, the signal rise and fall time are calculated and the signal charge is integrated numerically over a time window about the peak position.

For extracting the arrival time from the waveforms, we use the amplitude and rise time compensated (ARC) discrimination method which features a compensation for both amplitude and rise time [29]. The incoming pulse is duplicated into two identical components. One component is delayed. The other component is attenuated and inverted. The two components are then added together and the zero crossing time is computed as the timing of the incoming pulse. In this algorithm, the condition that must be met for the minimum rise time is $t_d \leq (1-f) t_{rise,\min}$, where $f$ is the attenuation factor, $t_d$ is the delay. The parameters are pulse shape dependent. For each particular detector, we experimentally adjust $f$ and $t_d$ to get the best result. Finally, slewing corrections for both amplitude and rise time are implemented, although these effects are negligible after a proper ARC discrimination.

Timing measurements are performed at various light intensities. The start time of the system is provided by the laser synchronization signal. The stop time is extracted from the MCP pulses through the ARC timing discrimination. In this paper, the arrival time is defined as the time difference between the MCP-PMTs and the laser synchronization signal. The time resolution is represented by the transit time spread (TTS) [1] that corresponds to the standard deviation of the arrival time distribution. The signals of the readout strips are recorded from both ends. The differential time between two ends of the same strip is used to determine the photon hit position.

## 6. Results and discussion

A number of 6 cm × 6 cm photodetectors have been produced and characterized at ANL. The QE of the photocathode has previously been demonstrated to reach a 15% at λ ~ 350 nm [24]. The test results presented are based on a typical Argonne prototype (see Fig. 3) with 20 μm MCP pore size.

### 6.1 Typical pulses

A typical pulse and its frequency components are shown in Fig. 6, obtained from the ANL device illuminated by a 405 nm laser light. The ANL device was operated at a very low light intensity with an average number of photoelectrons of about 1.4. For signal processing, we use the algorithm as described in Section 5. Fig. 6 (a) shows the raw signal waveform that is recorded by the oscilloscope at a sampling rate of 10 Gs/s for each channel. At the single photoelectron level, the low signal-to-noise ratio becomes a significant source to the time resolution. The noise level of the system is mainly determined by the design of the readout. The oscilloscope measurement in this paper presents a noise of about 2 – 3 mV. A signal filtering based on frequency domain analysis is performed to enhance the signal-to-noise ratio. Fig. 6 (b) shows the frequency spectrums of the MCP signal and noise. The signal is dominating up to a frequency of about 1 GHz; its frequency components higher than 1 GHz behave like noise. It is possible to filter out the high frequency noise by applying a low-pass frequency filter. We use a Butterworth filter and set the cutoff frequency at 900 MHz. Fig. 6 (d) shows a comparison between the raw and the filtered frequency spectra. The filtered signal can be obtained through an inverse FFT transform, as shown in Fig. 6 (c). The signal of the ANL device has a 5 ns pulse width over a threshold that is 10% of the pulse height. The rise time is about 700 ps, determined by the detector geometry and the voltage on each stack component.

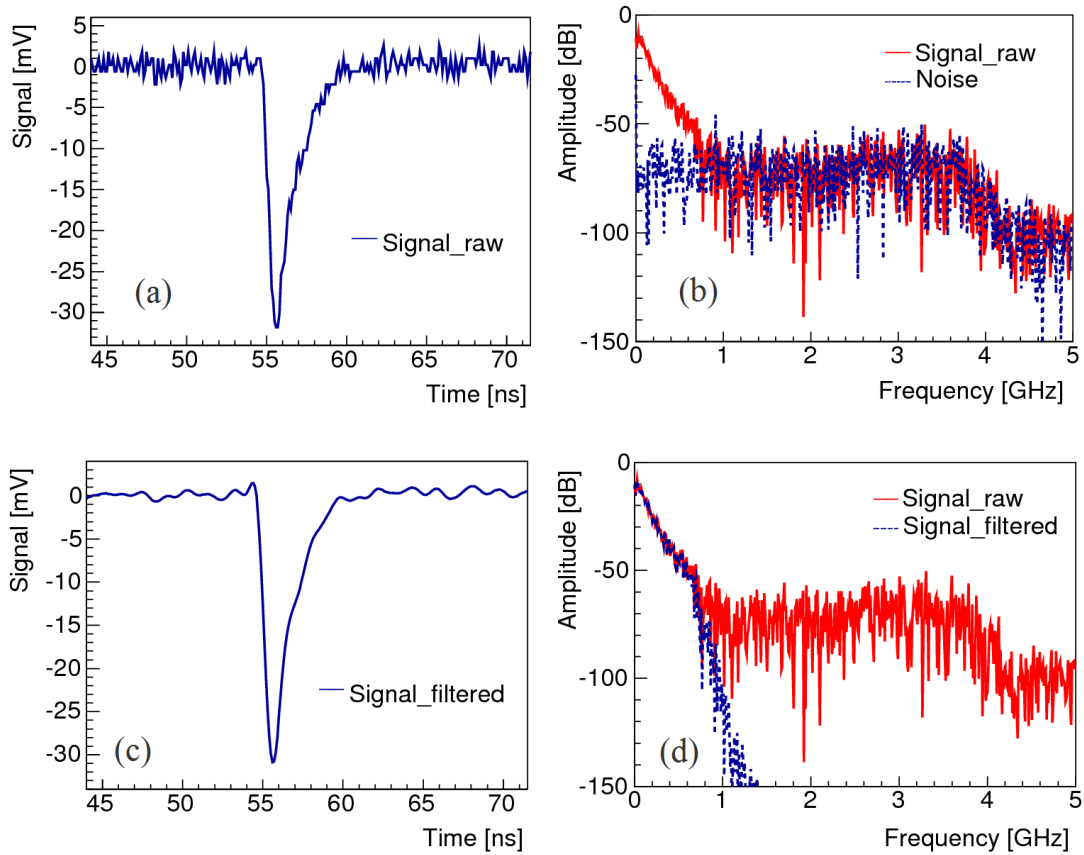

Fig. 6 Typical pulse shape and its frequency components. (a): Raw signal. (b): Frequency spectrum of the MCP noise (dash-blue) and signal (solid-red). (c) Filtered signal. (d): Comparison between the raw (solid-red) and filtered (dash-blue) frequency spectrums.

*6.2 Gain*

For the gain measurements, the light was attenuated to a very low level where the single PE regime was statistically reached. Fig. 7 shows the charge distribution of the device operated at 2560 V in single PE mode. The first peak corresponds to the pedestal. The broader second peak corresponds to the events with valid pulses. When the photodetector is working in the single PE mode as defined in Fig.5, most of the events in the second peak are from a single PE contribution and the multi PE contribution is well suppressed. The gain is measured directly by integrating the total charge in the second peak. Fig. 8 shows the typical gain of the device as a function of the applied HV. The gain ranges from $10^6$ to $10^7$ and a saturation effect is observed for a gain higher than $10^7$.

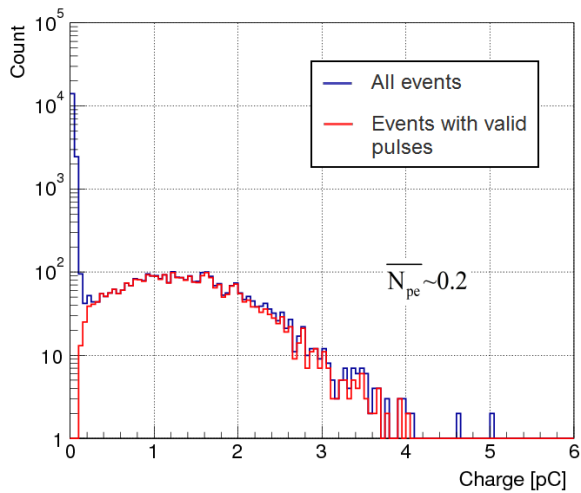

Fig. 7 Charge distribution of the ANL device in single PE mode.

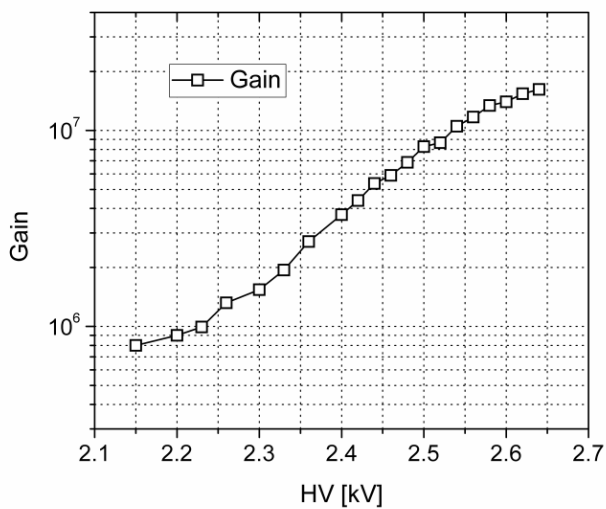

Fig. 8 Gain of the ANL device as a function of the applied HV.

### 6.3 Time response for single photoelectron

The laser controller[4] allows adjusting the optical output power. The laser data sheets provided by the manufacturer indicate that this product generates optical output of a single pulse when the power is lower than the nominal settings. If the power is at high settings, the optical pulse develops an oscillation shape: the first pulse is followed by one or two subsequent pulses with lower amplitude. We performed measurements with MCP-PMTs operating at different laser power settings. The goal was to measure the dependence of the timing distribution on the laser power.

---

[4] Model: C8988

Fig. 9 shows the timing distributions obtained with the Planacon and ANL device, at a high laser power and at low light intensity very close to the single PE level. Both detectors show a multi-peak sub-structure in the timing distribution, which confirms the existence of the laser oscillation sub-structure with a time interval of about 200 – 250 ps. Accordingly, the timing distribution is fit with four Gaussians, and the transit time spread resolution is inferred from the standard deviation of the first Gaussian. The second and the third Gaussians correspond to the optical oscillation pulses that arise at high laser power. We attribute the fourth Gaussian to the electron back-scattering that is dependent on the detector geometry. The ANL device shows a TTS of about 62 ps and the Planacon has a better resolution of about 40 ps. We attribute the difference to the MCP pore size and the light level. Compared to the Planacon (10 μm pore diameter), the ANL device is made of MCPs with a bigger pore size (20 μm pore diameter) and is operated at a slightly lower light level ($\overline{N}_{pe}$ ~0.35), resulting in a wider timing distribution.

The timing distributions at the nominal/low laser power are shown in Fig.10. We found that the laser optical oscillation pulses are well suppressed at this power. The timing distribution consists of a main prompt peak with a long tail. The main peak represents the TTS and the long tail is attributed to a combined effect from late laser pulses and photoelectron backscattering. It should be noted that in both Fig. 9 and Fig. 10, the laser jitter is included. A typical jitter of 70 ps FWHM is given for the laser at the nominal power setting, and the value will increase if the laser power is decreased. For most of the measurements in this paper, we chose to operate the detectors at the nominal power, where the optical oscillation is well suppressed.

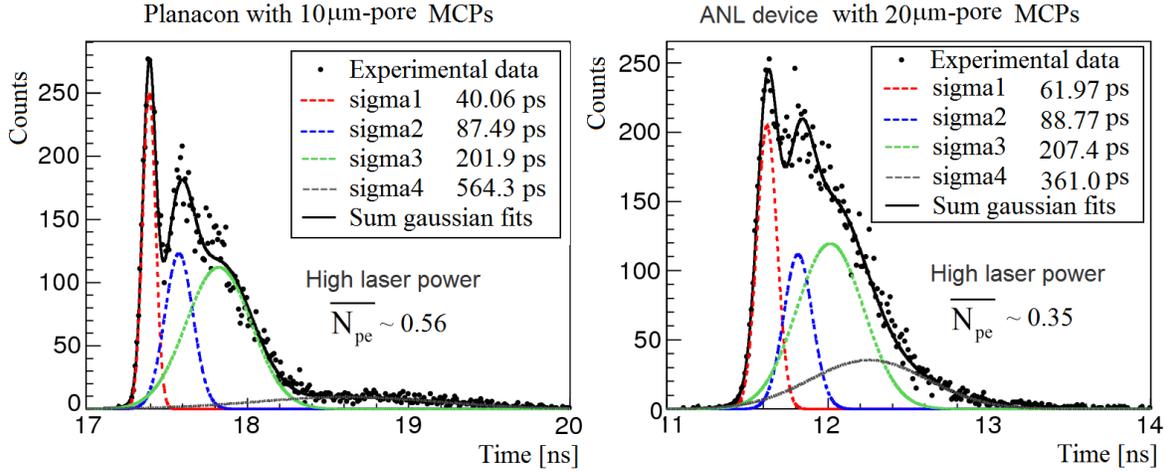

Fig. 9 Timing distribution for single PE light level, measured at high laser power. Left: Planacon with 10 μm - pore MCPs, HV = 2800 V. Right: ANL device with 20 μm - pore MCPs, HV = 2550 V.

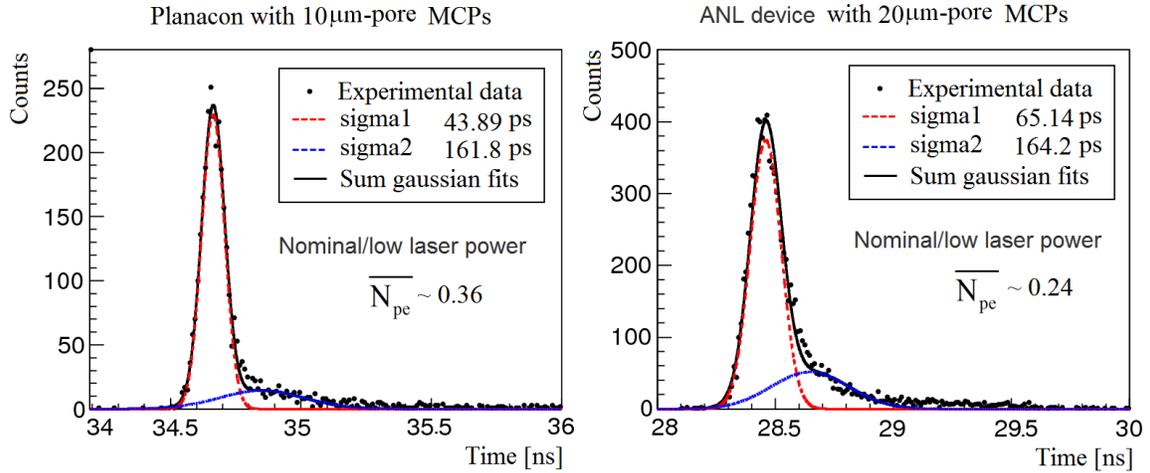

Fig. 10 Timing distribution for single PE light level, measured at the nominal laser power Left: Planacon with 10 μm - pore MCPs, HV = 2800 V. Right: ANL device with 20 μm - pore MCPs, HV = 2580 V.

At the single PE light level, timing measurements were performed for various HVs. The typical timing distribution obtained from single PE operation mode ($\overline{N_{pe}}$ <0.22) is presented in Fig. 11. The data points are fitted with polynomial functions to guide the eye. With the increase of the HV, the time resolution generally improves gradually and then reaches a plateau. The best system time resolution including all contributions from ($\sigma_{system}$, solid circles) is measured to be about 65 ps. The laser diode contributes $\sigma_{laser}$ ~30 ps to the TTS measurements in single photoelectron mode. The electronics contribution was measured to be $\sigma_{electronics}$ ~7 ps. By subtracting these contributions, we are able to extract the TTS of the MCP photodetector[5]. We estimate that the contribution from the MCP photodetector ($\sigma_{MCP-PMT}$, open circles) is about 57 ps.

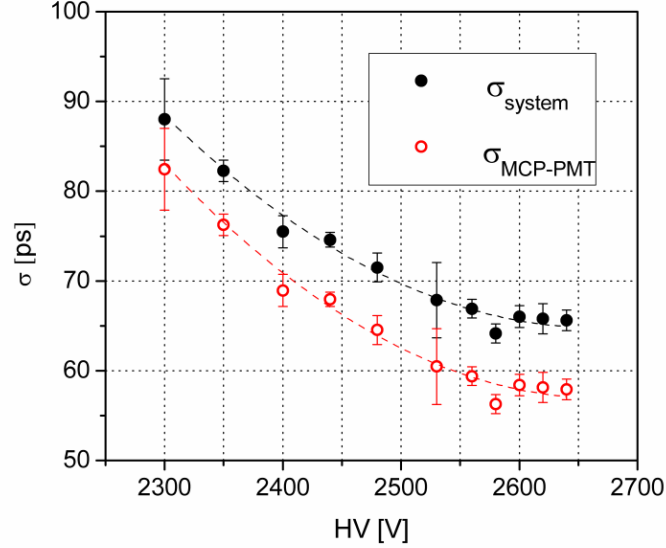

Fig. 11 Transit time spread (TTS) as a function of applied HV. Solid circles: system time resolution. Open circles: MCP-PMT time resolution with the laser and electronics contributions subtracted. The data points are fitted with a polynomial function to guide the eye.

*6.4 Time response for multi-photoelectrons*

We operated the photodetector at a HV of 2560 V where the gain is very high (about $10^7$), and performed measurements at various light intensities. Fig. 12 shows the measured time resolution as a function of the average number of photoelectrons that is estimated from a statistical analysis. Assuming that the number of photoelectrons $N_{pe}$ follows a Poisson distribution, the mean value $\bar{q}$ and the standard deviation $\sigma_q$ of the measured charge spectrum can be expressed as:

$$\bar{q} = eG\overline{N_{qe}}, \qquad \sigma_q = eG\sigma_{qe} = eG\sqrt{\overline{N_{qe}}} \quad (4)$$

where $e$ is the electron charge and $G$ is the gain. Therefore, the average number of photoelectrons $\overline{N_{qe}}$ is expressed as:

$$\overline{N_{pe}} = \left(\bar{q}/\sigma_q\right)^2 \quad (5)$$

In Fig. 12, the solid squares and the open circles represent the system and the MCP-PMT time resolutions[6], respectively. One can see that the laser diode contribution gets reduced with the increase of the light level. The

---

[5] For single photoelectron light level, the MCP-PMT contribution to the system resolution is:

$$\sigma_{MCP-PMT} = \sqrt{\sigma^2_{system} - \sigma^2_{laser} - \sigma^2_{electronics}}$$

time resolution of the 6 cm photodetector is initially about 60 ps, but improves at high light intensity, as expected. The best time resolution is about 15 ps, achieved for $\overline{N_{pe}} > 200$.

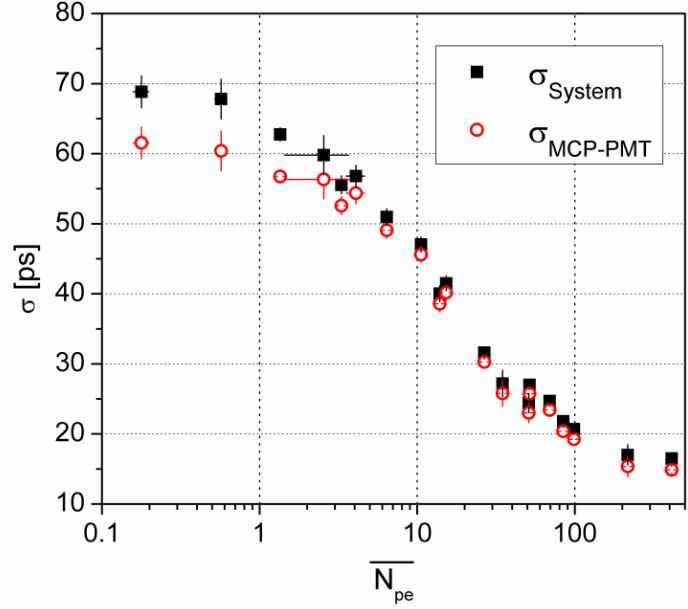

Fig. 12 Transit time spread (TTS) as a function of the average number of photoelectrons. Solid circles: system time resolution. Open circles: MCP-PMT time resolution with the laser and electronics contributions subtracted.

*6.5 Position resolution*

One advantage of the strip line anode is that the hit position can be reconstructed along the strip by comparing the time difference from two strip ends:

$$x_{position} = \frac{t_l - t_r}{2} v \qquad (6)$$

where $x_{position}$ represents the hit position, $t_l$ and $t_r$ are the signal arrival times of the left and right ends of a strip, $v$ is the signal transmission speed on the anode plate. To determine $v$, we performed a position scan along the strip at high light intensity. In Fig. 13, the time difference between two ends of a strip is plotted as a function of

---

[6] For multi-photoelectron light level, the MCP-PMT contribution to the system resolution is:

$$\sigma_{MCP-PMT} = \sqrt{\sigma_{system}^2 - \left(\frac{\sigma_{laser}}{\sqrt{N_{pe}}\big|_{amplitude>threshold}}\right)^2 - \sigma_{electronics}^2}$$

the position of the laser spot. The linear fit of the data points provides a signal transmission speed of about 184 µm/ps.

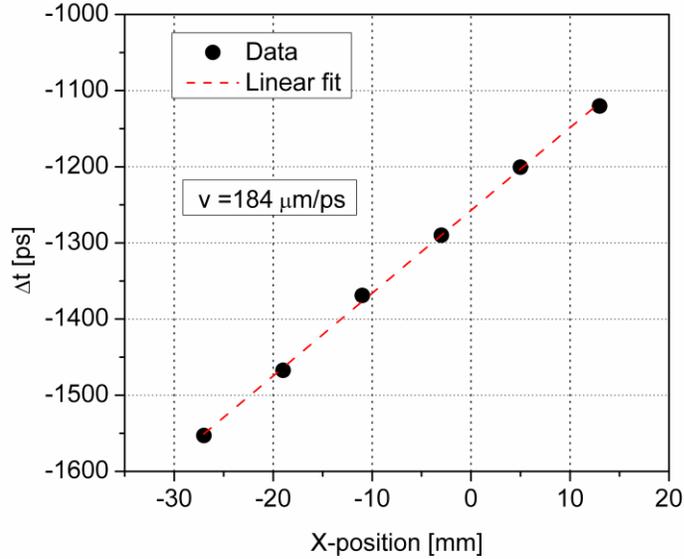

Fig. 13 Differential time between two ends of a strip as a function of the hit position along the strip.

Using the signal transmission speed, we are able to convert the differential time to hit position. As two strip ends see the same induced pulse, the MCP intrinsic time jitter resulting from the electron avalanche statistics is canceled in the differential time distribution. The position reconstructed by the differential time is therefore independent of the intrinsic MCP properties, but dependent on the quality of the pulse and readout electronics. Characteristics such as noise-to-signal ratio, bandwidth of the readout system, sampling rate, and signal transmission loss contribute to the overall position resolution. Fig.14 shows the distribution of the reconstructed position along the strip, for various light intensities. The distribution tends to be more Gaussian-like when the light intensity increases. As the position distribution obtained from this setup is a convolution of the beam size and the actual position resolution of the detector, the resolution shown there represents an upper bound.

We found that the position resolution of the detector is strongly dependent on the signal-to-noise ratio. The small signals form a wide distribution, while the large signals form a narrower distribution. The overall distribution is a sum of multiple Gaussians, resulting in a non-Gaussian shape. At low light intensity, the distribution deviates from the Gaussian shape because of the large relative variance of the pulse height. When one increases the light intensity, the position distribution becomes more Gaussian-like. We use a Gaussian fit (solid line) within $\pm 2\sigma$ to estimate the position resolution. The dashed line extends the Gaussian curve to $\pm 4\sigma$. The limiting resolution at very high $N_{pe}$ is found to be 0.75 mm, corresponding to a differential time resolution of 8.3 ps, compatible with the electronics contribution. The fluctuations at low $N_{pe}$ are mainly due to a high noise-to-signal ratio, which is of importance at low light intensity. Fig. 15 shows the correlation between the position resolution and the noise-to-signal (N/S) ratio. The left axis is the position resolution. The right axis is the corresponding differential time resolution. The photodetector at lower N/S ratio shows a better position resolution.

The position resolution at very low N/S ratio is limited by the electronics resolution and the beam size. The data points in the range 0.02<N/S<0.065 are fitted to a linear function. The extrapolation (dashed line) of the linear fit (solid line) shows a 0.2 mm and an equivalent 2.7 ps differential time resolution resolution at zero N/S ratio, which suggests that the position resolution can still be improved by increasing the MCP gain and optimizing the readout system. Similar observations were discussed previously for the demountable photodetector [22].

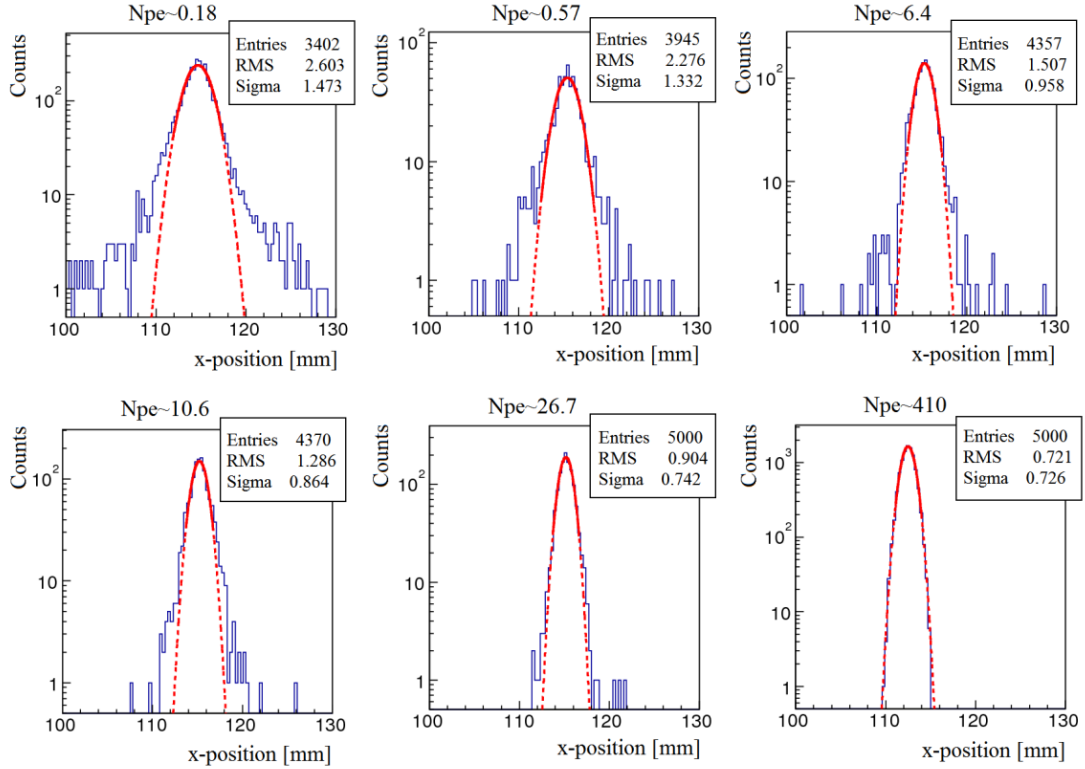

Fig. 14 Position distribution along the strip, for various light intensities. The solid line corresponds to a Gaussian curve fitted to within ±2σ to estimate the position resolution. The dashed line extends the Gaussian curve to ±4σ.

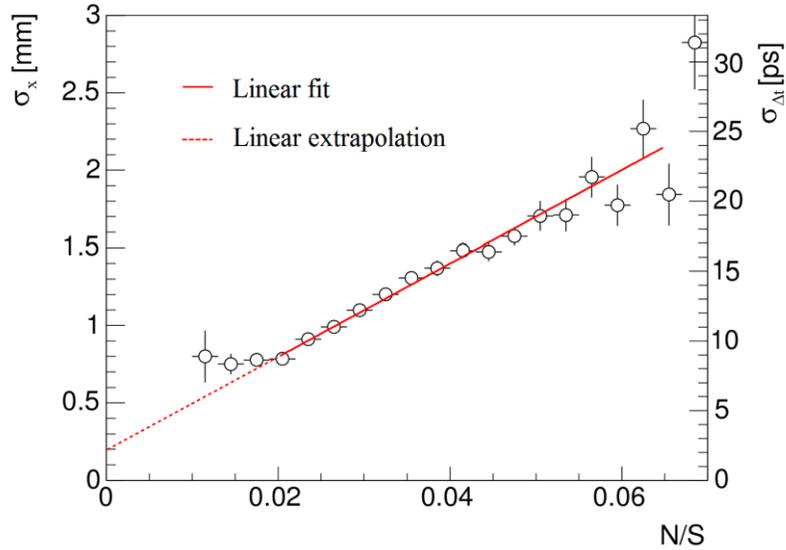

Fig. 15 Position resolution (upper bound) as a function of the noise-to-signal ratio. The left axis is the position resolution; The right axis is the corresponding differential time resolution. The solid line is a linear fit to the data in the range 0.02<N/S<0.065. The dashed line is a linear extrapolation, indicating the ideal resolution.

## 7. Conclusion

Argonne National Laboratory is currently focusing on the development of cost-effective, small form-factor (6 cm × 6 cm), MCP-based picosecond photodetectors with scalability to extend then to large area. Recently, a significant milestone of producing the first fully processed and hermetically sealed MCP-based photodetector has been reached at the small single tube processing system. A series of functional 6 cm × 6 cm prototypes have been produced and characterized. The R&D and production experience gained from fabricating these devices can be transferred to industry for large scale production. In this paper, we described the design of the photodetectors, the laser test facility, the data analysis technique, and the key performance parameters of the photodetectors.

For characterizing the performance of the photodetectors, we built a pulsed laser test facility. We have quantitatively shown how the single photoelectron light level can be reached without a detailed calibration of the light intensity. Using a 6 cm × 6 cm photodetector made of MCPs with 20 μm pores, we have achieved ~60 ps transit time spread resolution for single photoelectrons and ~15 ps for multi-photoelectrons. The upper bound on the position resolution is measured to be about 0.8 mm, including the beam size and the electronics contributions. Because of the limitation of the resistor chain design of the high voltage divider, all the results presented in this paper are obtained without optimization of the MCP high voltages. At the moment we are pursuing an alternate design, which will allow us to individually control and fine tune the voltage of each stack component to optimize the performance. We are also working towards an improvement of the tube processing system as well. Further studies, such as gain uniformity, rate capability, magnetic field effect, lifetime and properties at cryogenic temperature, will be pursued in the near future. Work on the QE optimization is also underway. The final goal is to reach an average 20% QE with uniformity better than 10%. Results will be published elsewhere.

## Acknowledgments

We would like to thank Ronald Kmak (ANL) for the design of the vacuum chamber. We also thank Joe Gregar (ANL) of the Argonne glass shop, for his talent work on the frit seal. We are deeply grateful to Matthew Wetstein (University of Chicago) and Bernhard Adams (ANL) for their support and advice on detector testing. Work at ANL was supported by the U.S. Department of Energy, Office of Science, Office of Basic Energy Sciences and Office of High Energy Physics under contract DE-AC02-06CH11357. Use of the Center for Nano scale Materials was supported by the U.S. Department of Energy, Office of Science, Office of Basic Energy Sciences, under Contract No. DE-AC02-06CH11357.